\documentclass{JHEP3}

\usepackage{epsfig}

\def\beqa{\begin{eqnarray}}
\def\eeqa{\end{eqnarray}}
\def\beq{\begin{equation}}
\def\eeq{\end{equation}}

\title{Caustic Formation in Tachyon Effective Field Theories}

\author{Neil Barnaby \\ McGill University, 3600 University St.\
Montr\'eal, Qu\'ebec H3A 2T8, Canada \\
E-mail: \email{barnaby@physics.mcgill.ca}}

\preprint{McGILL-14-04}

\abstract{
Certain configurations of D-branes, for example wrong dimensional branes or the brane-antibrane system,
are unstable to decay.  This instability is described by the appearance of a tachyonic mode in the 
spectrum of open strings ending on the brane(s).  The decay of these unstable systems is described by
the rolling of the tachyon field from the unstable maximum to the minimum of its potential.  We analytically
study the dynamics of the inhomogeneous tachyon field as it rolls towards the true vacuum of the theory
in the context of several different tachyon effective actions.
We find that the vacuum dynamics of these theories is remarkably similar and in particular we show that in 
all cases the tachyon field forms caustics where second and higher derivatives of the field blow up.  
The formation of caustics signals a pathology in the evolution since each of the effective actions 
considered is not reliable in the vicinity of a caustic.  We speculate that the formation of caustics is an 
artifact of truncating the tachyon action, which should contain all orders of derivatives acting on the 
field, to a finite number of derivatives.  Finally, we consider inhomogeneous solutions in
$p$-adic string theory, a toy model of the bosonic tachyon which contains derivatives of all orders acting 
on the field.  For a large class of initial conditions we conclusively show that the evolution is well 
behaved in this case.  It is unclear if these caustics are a genuine prediction of string theory or not.
}


\begin{document}

\section{Introduction}

Recently effective actions describing the open string tachyon have recieved considerable attention in the 
literature.  These effective actions are interesting to study since they permit a relatively simple
formulation of the complicated open string dynamics in terms of classical field theories.
In particular, the use of effective actions has triggered significant interest in the possible role of the
tachyon in cosmology.  For example, there have been numerous attempts to make use of the tachyon as either
the inflaton or as quintessence \cite{TachyonCosmology} (see \cite{Review} for reviews of tachyon cosmology).
Constraints and shortcomings of these  scenarios  have been discussed in \cite{Problems}. 

Perhaps the most 
promising application of the tachyon to cosmology comes from D-brane inflation \cite{DbraneInflation}.
In the context of D-brane inflation, the annihilation of a brane-antibrane pair at the endpoint of 
inflation is described by the rolling
of the tachyon, $T$, from the unstable maximum $T=0$ to the degenerate vacua of the theory $T=\pm \infty$.  
The tachyon 
field may roll to the same vacuum everywhere in the space or topological defects may form through the 
Kibble mechanism \cite{KibbleMechanism}.  The latter possibility corresponds to the formation of lower
dimensional branes at points where $T=0$.  In the regions between the descendant branes the tachyon field 
asymptotically rolls towards the true vacuum of the theory.  The formation of lower dimensional branes is of
some phenomenological importance since the resulting cosmic-string-like 
or higher dimensional defects could be observable remnants of inflation \cite{KibbleMechanism}.
In \cite{Reheating1,Reheating2} a more radical idea was explored: it was proposed that our own observable 
universe could be such a defect in 
the higher dimensional spacetime predicted by string theory.  In this scenario the coupling of the time 
dependent tachyon condensate to gauge particles  which will be localized on the descendant brane can provide
an efficient mechanism for reheating at the end of brane-antibrane inflation.

Applications of the tachyon to cosmology generically involve the tachyon field rolling from the unstable
maximum of its potential to the true vacuum of the theory, at least in some region of the space, so it
is of some interest to study the dynamics in this regime.  Often
only homogeneous tachyon profiles are considered, though a consistent study of the tachyon field in cosmology
must include spatial inhomogeneities.  In the context of the formation of lower dimensional branes at the
endpoint of D-brane inflation spatial inhomogeneity is unavoidable since the field must cross $T=0$ at one or
more points in the space.  However, even in the absence of topological defects one expects
that spatial inhomogeneities will be generated starting from a homogeneous profile by vacuum 
fluctuations of the tachyon field.

The full tachyon
action contains an infinite number of derivatives acting on the field, though the action 
is not known explicitly to all orders in derivatives.  Rather, numerous effective actions describing the 
tachyon have been proposed in the literature on different grounds and have been shown to reproduce various 
nontrivial aspects of the full dynamics expected from string theory.  
Most studies of the tachyon dynamics
in terms of effective actions have considered actions involving only first derivatives of the field and it 
is not clear how higher derivative corrections may alter the dynamics.  Indeed, studies of $p$-adic string 
theory suggest that there may be fundamental differences between dynamical equations with an infinite number
of derivatives and those with a large finite number of derivatives.  In \cite{Zwiebach:p-adic} it was shown 
that the initial value problem is qualitatively different in these two cases.

In this paper we study analytically the dynamics of the inhomogeneous tachyon field as it rolls towards
the vacuum in the context of several different effective field theories.  Our 
interest is motivated by \cite{Caustics} in which inhomogeneous vacuum solutions were discussed using the 
popular tachyon Dirac-Born-Infeld (DBI) action.  The authors of 
\cite{Caustics} found that in general the inhomogeneous solutions of the DBI action formed caustics where the
second and higher derivatives of the tachyon field blow up at some point.  In the vicinity of these caustics
the tachyon DBI action is not reliable since the higher derivative corrections which have been ignored
will become important.
It is thus not known if the formation of caustics is a genuine 
prediction of string theory or an artifact of the DBI action for the tachyon.  To shed some light on this 
question it is interesting to study inhomogeneous vacuum solutions of different tachyon effective 
actions and look for caustics or similar singularities.

The finite time formation of caustics found in \cite{Caustics} must be distinguished from the finite time
divergences in the first derivatives of the tachyon field at points where $T=0$ \cite{Reheating2,RealTime}
which corresponds to the formation of topological defects.  In both cases the divergences in the derivatives
of the tachyon field lead to divergences in the energy density.  In the case of topological defect formation,
however, this finite time divergence in the energy density has the form of a delta function and hence 
leads to finite total energy \cite{Reheating2}.
We point out that similar behaviour was found in \cite{Sbranes} where the decaying tachyon on the 
supergravity background corresponding to space-like branes in string theory was studied.  In this case the 
first derivative of the tachyon field and the energy density of the tachyon matter both divergence in finite
time producing a space-like curvature singularity in the metric.

The organization of this paper is as follows.
In section \ref{FirstDerivative} we consider three tachyon effective actions, all of which contain only 
first derivatives 
of the field and show that all three display surprisingly similar dynamics near the true vacuum of the 
theory.  In section \ref{CausticFormation} we briefly review how this universal vacuum structure leads to 
the formation of 
caustics.  In section \ref{SecondDerivative} we consider the vacuum dynamics of the tachyon in the context 
of yet another effective action which contains terms with second derivatives of the field and show that the 
problem of caustic formation is not ameliorated.  In section \ref{InfiniteDerivatives} we briefly 
review the bosonic tachyon action of $p$-adic string theory and study small spatial inhomogeneities about a 
time dependent solution which rolls towards the minimum of the potential (which is unbounded from below).
We find that for a large class of initial data our perturbative expansion is reliable throughout the 
evolution and the solutions are well behaved.  Finally we conclude and discuss the difficulty of 
interpreting caustic formation physically.

\section{First Derivative Tachyon Effective Actions and the Eikonal Equation \label{FirstDerivative}}

For simplicity we restrict ourselves to studying real tachyon fields in $1+1$-dimensional Minkowski space 
with metric $\eta_{\mu\nu}dx^{\mu}dx^{\nu} = -dt^2 + dx^2$.  In the context of brane annihilation the rolling
of the tachyon in $1+1$ dimensions from the unstable maximum to the ground state describes the decay of an 
unstable D$p$-brane to either the vacuum (in the case that the tachyon rolls to the same vacuum everywhere)
or to a brane of codimension one (in the case that kinks form).  We work in units where 
$\alpha'=2$.

\subsection{The Tachyon Dirac-Born-Infeld Action}
\label{TDBI}

We first consider the tachyon Dirac-Born-Infeld action \cite{DBIAction,Sen:TachyonMatter,Sen:KinkAndVortex}

\begin{equation}
\label{DBI}
  \mathcal{L} = -U(T) \sqrt{1+\partial_{\mu} T \partial^{\mu} T}
\end{equation}
with potential
\begin{equation}
\label{U(T)}
  U(T) = \frac{\tau_p}{\cosh(T/2)}
\end{equation}
where $\tau_p$ is the D$p$-brane tension.
Solutions of this theory have been widely studied in the literature 
\cite{Reheating1,Reheating2,Caustics,Sen:KinkAndVortex,DBIStudies,VacuumDBI}.  This action can be obtained 
from string theory 
in some limit \cite{DBIderive} and has been shown to reproduce numerous nontrivial aspects of the full
string theory dynamics \cite{Sen:KinkAndVortex,DBIproperties}.

The action (\ref{DBI}) should be thought of as describing tachyon profiles which are ``close''
\footnote{That is to say taking $T_{+}$, $T_{-}$ to be slowly varying functions of $x^{\mu}$.} 
to the homogeneous rolling tachyon solution

\begin{equation}
\label{rollingsolution}
  T(t)  = 2 \sinh^{-1} \left( T_{+} e^{t / 2} + T_{-} e^{- t / 2} \right).
\end{equation}

We are interested in the dynamics of (\ref{DBI}) close to the vacuum $T \rightarrow \pm \infty$, 
$U(T) \rightarrow 0$.  The vacuum structure of this theory has been studied analytically in 
\cite{Sen:TachyonMatter,VacuumDBI} which we briefly review.   The vacuum structure is most easily described 
in the Hamiltonian formalism since the Lagrangian does not survive the limit $U(T) \rightarrow 0$, but the 
Hamiltonian does.  Defining the momentum conjugate to $T$ as $\Pi = \partial \mathcal{L} / \partial \dot{T}$
\footnote{Here and throughout this paper $\dot{T} = \partial_0 T  = \partial_t T$ and
$T' = \partial_1 T = \partial_x T$.} 
the Hamiltonian is given by 
\[
   H = \int d^2 x \, \mathcal{H}, \hspace{5mm} 
   \mathcal{H} = \sqrt{\Pi^2 + U(T)^2 } \sqrt{ 1 + T'^2 }.
\]
In the limit as $U(T) \rightarrow 0$ one finds
\begin{equation}
\label{hamiltonVac}
  \mathcal{H} = \Pi \sqrt{ 1 + T'^2 }.
\end{equation}
In the vacuum Hamilton's equations of motion are
\begin{equation}
\label{hamilton1}
  \dot{T} = \sqrt{ 1 + T'^2 }
\end{equation}
and

\begin{equation}
\label{hamilton2}
  \dot{\Pi} = \partial_x \left( \Pi \frac{T'}{ 1 + T'^2 }  \right)
\end{equation}
for $\Pi>0$.  Equation (\ref{hamilton1}) is precisely the eikonal equation
\begin{equation}
\label{eikonal}
   \partial_{\mu} T \partial^{\mu} T +1 = 0.
\end{equation}
It was observed numerically in \cite{Caustics} that the general solutions of (\ref{DBI}) tend towards the 
first order equation (\ref{eikonal}).

\subsection{The Boundary String Field Theory Action}

The action

\begin{equation}
\label{BSFT}
  \mathcal{L} = -V(T) F \left(  \partial_{\mu} T \partial^{\mu} T  \right)
\end{equation}
where the potential is

\begin{equation}
\label{potential}
  V(T) = \sqrt{2} \tau_p \exp \left( -\frac{T^2}{4} \right)
\end{equation}
and the kinetic term is

\begin{equation}
\label{F(z)}
  F(z) = \frac{4^z z \Gamma(z)^2}{2 \Gamma(2z)}
\end{equation}
can be derived from boundary string field theory (BSFT) assuming a linear profile $T = a + u_{\mu} x^{\mu}$,
 and summed
to all orders in $u_{\mu}$ \cite{BSFTAction}.  This action should therefore be considered reliable only
when describing profiles where second and higher derivatives of the tachyon field are small.

The dynamics of the action (\ref{BSFT}) have been discussed in \cite{Ishida:RollingDownToDbrane} which
we briefly review.  Before attempting to describe the vacuum dynamics of the 
theory (\ref{BSFT}) we discuss some limiting behaviour of the function $F(z)$.

At small $z$ the function $F(z)$ has a Taylor expansion

\[
  F(z) \cong 1 + \ln(4) z + \cdots
\]
and at large positive $z$ the function $F(z)$ has the behaviour

\begin{equation}
\label{largez}
  F(z) \rightarrow \sqrt{\pi z} \hspace{5mm} \mbox{as $z \rightarrow \infty$}.
\end{equation}
It is also noteworthy that $F(z)$ is singular at $z=-n$ ($n=1,2,\cdots$).  Of particular interest to the 
ensuing analysis will be the limiting behaviour of $F(z)$ near the singular point $z=-1$

\begin{equation}
\label{z=-1}
  F(z) \cong \frac{-1}{2(z+1)}, \hspace{5mm} F'(z) \cong \frac{1}{2(z+1)^2}, 
  \hspace{5mm} F''(z) \cong\frac{-1}{(z+1)^3}.
\end{equation}

We now proceed to study the dynamics of the action (\ref{BSFT}).
The equation of motion which follows from (\ref{BSFT}) is

\begin{equation}
\label{BSFTeom}
  \partial^{\mu} T \partial_{\mu} z \, F''(z) + \partial^{\mu} \partial_{\mu} T \, F'(z) 
   - \frac{1}{2} T \, z F'(z) + \frac{1}{4} T \, F(z) = 0 
\end{equation}
where $z = \partial_{\mu} T \partial^{\mu} T $.  We will assume initial conditions such that $z \approx 0$
initially.  Furthermore, we assume that z does not cross the singularity at $z=-1$ \footnote{This is not
a particularly restrictive assumption since if $z$ did cross the singularity it would lead to infinite
action.  Note that this does not exclude the possibility that $z \rightarrow -1$ as $V(T) \rightarrow 0$.}.

The homogeneous solutions of (\ref{BSFTeom}) have the asymptotic behaviour $T \rightarrow \pm t$ at
late times \cite{BSFThomogeneous}
so we expect that the tachyon will roll towards the vacuum $T \rightarrow \pm \infty$ as 
$t \rightarrow \infty$.  We consider first the case where $z$ does not approach $-1$ as 
$T \rightarrow \infty$.
In this case $F(z)$, $F'(z)$ and  $F''(z)$ are well-behaved and the leading contribution to (\ref{BSFTeom}) 
is

\[
  F(z) - 2 z F'(z) =0.
\]
It is easy to show that the quantity $ F(z) - 2 z F'(z)$ is greater than or equal to $0$ for all $z>-1$
and that $ F(z) - 2 z F'(z) \rightarrow 0$ as $z \rightarrow +\infty$ (see equation (\ref{largez})).  We 
conclude then that for solutions
where $z$ is increasing the equation of motion requires $z \rightarrow +\infty$ as 
$T \rightarrow \pm \infty$.

We now consider the possibility that $z \rightarrow -1$ at late times.  Near $z=-1$ the function $F(z)$ and
its derivatives are given by (\ref{z=-1}) and hence the leading contribution to (\ref{BSFTeom}) is

\begin{equation}
\label{eikonalleading}
  \partial_{\mu} T \partial^{\mu} z \, F''(z) - \frac{1}{2} T \, z F'(z) =0.
\end{equation}
This equation can be satisfied for profiles where $z \rightarrow -1$.  Since $z+1 \rightarrow 0$ as 
$T \rightarrow \pm \infty$ it is reasonable to assume that we can write $z+1=f(T)$.  This is consistent with
(\ref{eikonalleading}) which implies
\[
  z+1 \sim V(T)^{1/2}.
\]

It is important to stress that, barring any asumptions on the form of the solution, the equation of motion
(\ref{BSFTeom}) requires that  $\partial_{\mu} T \partial^{\mu} T$ tends to either $+\infty$ or $-1$ as
$T \rightarrow \pm \infty$.  The former case describes tachyon kinks of the form $T(x) = x / \epsilon$
to be understood in the limit that $\epsilon \rightarrow 0$.  The latter case describes the 
rolling tachyon and corresponds to the vacuum structure we are interested in.  We conclude then that close
to the vacuum the tachyon field obeys the eikonal equation

\[
   \partial_{\mu} T \partial^{\mu} T +1 = 0.
\]

\subsection{The Lambert and Sachs Action}

The action

\begin{equation}
\label{Lambert}
  \mathcal{L} = -V(T) K( \partial_{\mu} T \partial^{\mu} T )
\end{equation}
where the potential is given by (\ref{potential}) and the kinetic term is

\begin{equation}
\label{K(z)}
  K(z) = e^{-z} + \sqrt{\pi z} \, \mathrm{erf} \left( \sqrt{z} \, \right)
\end{equation}
was proposed in \cite{LambertAction}.
This action is derived by requiring that the profile

\[
  T(x) = \chi \sin \left( \frac{x-x_0}{2} \right)
\]
be an exact solution to the equations of motion.  The residual freedom is fixed by demanding that 
$\mathcal{L} = \sqrt{2} \tau_p \exp (-T^2 / 4 )$ for constant profiles for agreement with 
(\ref{BSFT}).

Before proceeding to study the dynamics of (\ref{Lambert}) we consider the asymptotic behaviour of the
kinetic term (\ref{K(z)}).
For $z \geq 0$ the function $K(z)$ is remarkably similar to the BSFT function $F(z)$ given by (\ref{F(z)}).
Near $z=0$ the function $K(z)$ has a Taylor expansion

\[
  K(z) \cong 1 + z + \cdots.
\]
For large positive values of $z$ the function $K(z)$ has the aymptotic behaviour

\begin{equation}
\label{largepositivez}
  K(z) \rightarrow \sqrt{\pi z}, \hspace{1mm} K'(z) \rightarrow \frac{1}{2} \sqrt{\frac{\pi}{z}},
  \hspace{1mm} K''(z) \rightarrow - \frac{\sqrt{\pi}}{4 z^{3/2}}
\end{equation}
so that $K(z)$ tends to infinity slowly as $z \rightarrow +\infty$ while $K'(z)$, $K''(z)$ tend to zero.
Finally we consider the behaviour of $K(z)$ for large negative $z$.  Using the formula

\[
\sqrt{\pi} \, \mathrm{erf}(iy) \rightarrow i e^{y^2} 
\left( \frac{1}{y} + \frac{1}{2 y^3} + \mathcal{O}(y^{-5}) \right)
\]
for $y \rightarrow \infty$ it is easy to show that

\begin{equation}
\label{largez<0}
  K(z) \rightarrow - \frac{e^{|z|}}{2|z|}, \hspace{5mm} 
  K'(z) \rightarrow + \frac{e^{|z|}}{2|z|}, \hspace{5mm}
  K''(z) \rightarrow - \frac{e^{|z|}}{2|z|}
\end{equation}
for large negative $z$.  Note that though $K(z)$ has very similar behaviour to $F(z)$ and $\sqrt{1+z}$ for 
large positive $z$, these three kinetic terms have very different behaviour on $z<0$.

The equation of motion which follows from (\ref{Lambert}) is
\begin{equation}
\label{Lamberteom}
  \partial^{\mu} T \partial_{\mu} z \, K''(z) + \partial^{\mu} \partial_{\mu} T \, K'(z) 
   - \frac{1}{2} T \, z K'(z) + \frac{1}{4} T \, K(z) = 0 
\end{equation}
where $z=\partial_{\mu} T \partial^{\mu} T$.  By construction equation (\ref{Lamberteom})
has the homogeneous solution
\[
  T(t) = T_{+} e^{t / 2} + T_{-} e^{- t / 2}.
\]
Comparing this to the homogeneous solution (\ref{rollingsolution}) demonstrates that
a field redefinition is necessary to make
accurate comparisons between the two theories.  Notice that in the case of the homogeneous solution the
tachyon field rolls towards the vacuum $T \rightarrow \pm \infty$ at late times with 
$z \rightarrow -\infty$.

We proceed now to study the vacuum structure of the theory (\ref{Lambert}).  First we consider the 
possibility that $z$ tends to some finite positive value as $T \rightarrow \pm \infty$.  In this 
case the leading contribution to (\ref{Lamberteom}) is

\begin{equation}
\label{leading}
  T z K'(z) = \frac{1}{2} T K(z)
\end{equation}
or

\[
  0 = e^{-z}.
\]
This suggests that we require $z \rightarrow +\infty$ if $z>0$ and indeed, using the limiting behaviour 
(\ref{largepositivez}) the leading contribution to (\ref{Lamberteom}) for $T \rightarrow \pm \infty$
with $z \rightarrow +\infty$ is satisfied.  

We consider now possibility that $z$ approaches some finite negative value as 
$T \rightarrow \pm \infty$.
Assuming $z \not= -\infty$ then $K(z)$, $K'(z)$ and $K''(z)$ are all well behaved and the leading 
contribution 
to (\ref{Lamberteom}) is still (\ref{leading}) or equivalently $0 = e^{|z|}$ which has no solution.

Finally we consider the possibility that $z \rightarrow -\infty$ as $T \rightarrow \pm \infty$.  
In this
case the limiting behaviour (\ref{largez<0}) is applicable and $K(z) \sim -K'(z) \sim K''(z)$ so that the 
leading contribution to (\ref{Lamberteom}) is

\begin{equation}
\label{leading2}
  2 \partial_{\mu} T \partial^{\mu} z + T z =0.
\end{equation}
Since $z \rightarrow -\infty$ as $T \rightarrow \pm \infty$ it is reasonable to assume 
$z=f(T)$ 
in this regime. This assumption is consistent with (\ref{leading2}) and we find 
$f(T) = -T^2$ up to an arbitrary additive constant which is irrelevant as
$T^2 \rightarrow \infty$.  One may verify that given
these assumptions the leading behaviour (\ref{leading2}) is order $T^3$ while the terms we have 
disregarded in (\ref{Lamberteom}) are order $T$ so that this series of approximations is 
self-consistent.  We conclude that for $z<0$ the equation of motion requires that the tachyon field obeys

\[
  \partial_{\mu} T \partial^{\mu} T + T^2 =0.
\]
Defining $T = \exp(\tilde{T})$ we find that $\tilde{T}$ obeys the eikonal equation

\[
   \partial_{\mu} \tilde{T} \partial^{\mu} \tilde{T} + 1 =0.
\]
We stress that the equation of motion (\ref{Lamberteom}) requires either $z \rightarrow +\infty$
or $z \rightarrow -\infty$ as $T \rightarrow \pm \infty$.  The former case describes tachyon
kink solutions while the latter case describes the rolling tachyon and corresponds to the vacuum structure 
we are interested in.

\section{Caustic Formation\label{CausticFormation}}

We have found that all three actions (\ref{DBI},\ref{BSFT},\ref{Lambert}) predict that the tachyon field is
described by
the eikonal equation (\ref{eikonal}) close to the vacuum.  Inhomogeneous solutions of (\ref{eikonal}) were
found in \cite{Caustics} for arbitrary Cauchy data by the method of characteristics.
The solution is given along a set of characteristic curves which,
in $1+1$ dimensions, are defined by

\begin{equation}
\label{curve}
  x(q,t) = q - \frac{T_i'(q)}{\sqrt{1+T_i'(q)^2}}t
\end{equation}
where the parameter $q$ defines the initial position of the curve
on the $x$ axis such that $x(q,t=0)=q$ and 
$T_i(x)$ is the Cauchy data at time $t=0$.  The parameter $q$ should be thought of as labelling the curves.  
The value of the tachyon field along a given characteristic curve is

\begin{equation}
\label{T}
  T(q,t) = T_i(q) + \frac{t}{\sqrt{1+T_i'(q)^2}}.
\end{equation}
It is also worth noting that the derivatives of the field are constant along the curves $T'(q,t) = T_i'(q)$
and $\dot{T}(q,t) = \sqrt{1+T_i'(q)^2}$.  From (\ref{curve}) it is clear that in general
for initial data 
where $T_i'' \not= 0$ there will be curves originating from different initial points on the $x$ axis which
will cross in some finite time.  At points where the characteristic curves intersect the field becomes 
multi-valued since evolving (\ref{T}) along two different curves leads to two different values of $T$ at the
point of intersection.  This corresponds to the formation of caustics and signals a pathology in the 
evolution.  To illustrate the problem, consider $T''$ along a characteristic curve

\[
  T''(q,t) = \frac{T_i''(q)}{ 1 - \frac{T_i''(q)}{(1+T_i'(q)^2)^{3/2}}t }.
\]
At some time $t_c$ and point $q_c$ where $T_i''(q_c) > 0$
the denominator vanishes and the second derivative blows up.  For a given
$q_c$ the caustics form at time

\[
  t_c = \frac{(1+T_i'(q_c)^2)^{3/2}}{T_i''(q_c)}.
\]
In each of the three effective actions considered above (\ref{DBI},\ref{BSFT},\ref{Lambert}) this caustic
formation signals a breakdown of the theory since terms in the action involving second and higher derivatives
of the field which have been neglected will become important.

\section{Vacuum Dynamics in a Higher Derivative Action \label{SecondDerivative}}

In light of the analysis of the preceeding section it is tempting to speculate that the caustic
formation described above is an artifact of the derivative truncation which leads to 
(\ref{DBI},\ref{BSFT},\ref{Lambert}) since these effective descriptions break down in the vicinity of a 
caustic.
We would like to reconsider the effective description of the tachyon near the vacuum in the context of an 
action which does contain terms with second derivatives of the field.  
One might consider attempting to generalize the action (\ref{BSFT}) to a profile of the
form $T = a + u_{\mu} x^{\mu} + v_{\mu\nu} x^{\mu} x^{\nu}$ to obtain an action which is valid for profiles
where $\partial_{\mu} T \not= 0$ and $\partial_{\mu} \partial_{\nu} T \not= 0$ without constraint on the size
of the first and second derivatives.  However, deriving such an action (following \cite{BSFTAction}) 
would involve performing a path integral which is no longer Gaussian.
On the other hand, the superstring tachyon effective action has
been calculated in an expansion in small momenta (derivatives) around a constant profile up to six orders
in derivatives ($\mathcal{O}(\partial^6)$) in \cite{Laidlaw:EffectiveAction}.  The action, in units where 
$\alpha'=2$ and truncated to fourth order in derivatives, is
\begin{eqnarray}
\label{Laidlaw}
  \mathcal{L} = -V(T)\hspace{1mm} ( &1 + \ln(4) \partial_{\mu} T \partial^{\mu} T 
                +\left(  \frac{1}{2}(\ln(4))^4  - \zeta(2) \right) (\partial_{\mu} T \partial^{\mu} T)^2 
          +   \zeta(2) T \partial_{\mu} \partial_{\nu} T \partial^{\mu} T \partial^{\nu} T  & \nonumber \\
 & + \left[ (2 \zeta(2) - 8(\ln(2))^2 ) + \frac{\zeta(2)}{2} T^2  \right]  
\partial_{\mu} \partial_{\nu} T  \partial^{\mu} \partial^{\nu} T& 
\!\!\!\!\!\!\!\!\!\!\!\!\!\!\!\!\!\!\!\!\!\!\!\!\!\!\!\!\!\!\!\!\!\!\!) 
\end{eqnarray}
where $\zeta(z)$ is the Riemann zeta function.
The action (\ref{Laidlaw}) is valid to all orders in $T$ but for small derivatives and hence naively does 
not seem
applicable to studying the vacuum dynamics where, the analysis of the preceeding sections would suggest, 
$\partial_{\mu} T \partial^{\mu} T$ is order unity or larger.  However, one might expect the vacuum structure
of the theory to be significantly modified due to the presence of terms like 
$T^2 \partial_{\mu} \partial_{\nu} T  \partial^{\mu} \partial^{\nu} T $ in the action (\ref{Laidlaw}) which
are vanishing on linear profiles but are large in the vacuum for nonlinear profiles.  One might hope that 
the vacuum structure of the theory (\ref{Laidlaw}) is such that when one considers Cauchy data where the
derivatives of $T$ are small but nonzero, then the derivatives stay small throughout the evolution and the
approximations which lead to (\ref{Laidlaw}) remain self consistent.  This turns out not to be the case, as 
we will show.

The equation of motion which follows from (\ref{Laidlaw}) is cumbersome and we do not write it out here.
In the vacuum, as $T \rightarrow \pm \infty$, assuming that the derivatives of the field are well
behaved
\footnote{If we drop this assumption then (\ref{Laidlaw}) ceases to be a reasonable description of the 
dynamics.}
the leading contribution to the equation of motion is

\begin{equation}
\label{leadinghigher}
  \partial^{\mu} \partial^{\nu} T \partial_{\mu} T \partial_{\nu} T = 0
\end{equation}
where the subleading terms are of order $T^{-1}$ and smaller.

We note that (\ref{leadinghigher}) may be re-written as
\[
  \partial_{\mu} T \partial^{\mu} \left( \partial_{\nu} T \partial^{\nu} T \right) = 0
\]
so that the solution set of (\ref{leadinghigher}) contains the solutions of the first order equation
\[
 \partial_{\nu} T \partial^{\nu} T = \kappa 
\]
where $\kappa$ is a constant which may be set to $+1$, $-1$ or $0$ by rescaling $T$.
The case $\kappa=-1$ corresponds to the eikonal equation and we conclude that 
(\ref{leadinghigher}) does admit solutions with caustic formation in the case of Cauchy data such that 
$\dot{T}_i(x)^2 = 1 + T_i'(x)^2$.  However, equation (\ref{leadinghigher}) is second order and of course it 
is not necessary to constrain the Cauchy data in this manner.  For completeness we discuss the construction
of more general solutions of (\ref{leadinghigher}) in the appendix.

It is beyond the scope of this paper to perform a comprehensive analysis of the derivative singularities
admitted by (\ref{leadinghigher}). We have shown an explicit reduction of (\ref{leadinghigher}) to the
eikonal equation, which exhibits caustics.  As discussed in the appendix, we believe more general solutions
of (\ref{leadinghigher}) exhibit similar pathologies.  For our purposes, this is sufficient to demonstrate 
that the higher derivative action (\ref{Laidlaw}) does not ameliorate the problem of caustic formation.  In 
light of these results, it is tempting to speculate that it is necessary to use an action which contains all
orders of derivatives acting on the field to obtain a fully consistent description of the dynamics of the 
tachyon near the vacuum.

As an illustrative example we will discuss inhomogeneous solutions in a $p$-adic string theory, a toy theory
of the bosonic string tachyon which contains an infinite number of derivatives acting on the field which is 
known explicitly to all orders in derivatives. 

\section{Inhomogeneous Solutions in $p$-adic String Theory \label{InfiniteDerivatives}}

The action of $p$-adic string theory is \cite{Witten}

\begin{equation} 
\label{p-adic}
  S = \frac{p^2}{g^2(p-1)} \int d^D x \left[ 
  -\frac{1}{2}\phi \, p^{-\frac{1}{2} \partial_{\mu}\partial^{\mu} } \phi  + \frac{1}{p+1} \phi^{p+1} \right]
\end{equation}
where $\phi$ is the open string tachyon, $g$ is the open string coupling constant and, though the action was
derived for $p$ a prime number,
it appears that $p$ can be continued to any positive integer (the action makes sense even in
the limit $p \rightarrow 1$ \cite{p=1}).  The differential operator 
$p^{-\frac{1}{2} \partial_{\mu}\partial^{\mu} }$ is to be understood as the series expansion
\[
  p^{-\frac{1}{2} \partial_{\mu}\partial^{\mu} } 
= \exp \left( -\frac{1}{2} \ln(p) \, \partial_{\mu}\partial^{\mu}  \right) 
= \sum_{n=0}^{\infty} 
\left(-\frac{1}{2}\ln(p) \right)^n \frac{1}{n!} \left( \partial_{\mu}\partial^{\mu} \right)^n.
\]
The equation of motion which follows from (\ref{p-adic}) is
\begin{equation}
\label{p-adicEOM}
    p^{-\frac{1}{2} \partial_{\mu}\partial^{\mu} } \phi = \phi^p
\end{equation}
and the potential is
\begin{equation}
\label{p-adicPotential}
  V ( \phi ) = \frac{p^2}{g^2(p-1)} \left(  \frac{1}{2} \phi^2  - \frac{1}{p+1}\phi^{p+1} \right).
\end{equation}

\DOUBLEFIGURE[ht]{padic1.eps,width=2in,height=1.5in}{padic2.eps,width=2in,height=1.5in}{Plot of the potential $V(\phi)$ for even $p$.\label{padicFIG1}}{Plot of the potential $V(\phi)$ for odd $p$.\label{padicFIG2}}

The cases of odd and even $p$ are qualitatively different.  For odd $p$ the potential is an even function of
$\phi$ and for even $p$ the potential is an odd function of $\phi$.  In both cases the potential is 
unbounded from below.  In the case of even $p$ the perturbative vacuum is at $\phi=1$ and in the case of odd
$p$ there is an equivalent false vacuum at $\phi=-1$.  In both cases the true vacuum of the theory is at
$\phi = 0$.  Figures \ref{padicFIG1} and \ref{padicFIG2} show plots of the potential (\ref{p-adicPotential})
for the cases of even and odd $p$ respectively.

We note that the action (\ref{p-adic}) is intended to describe the bosonic string tachyon, rather than the
superstring tachyon which we have considered in our previous analysis 
(\ref{DBI},\ref{BSFT},\ref{Lambert},\ref{Laidlaw}).  Furthermore, (\ref{p-adic}) is a toy model which is
only expected to qualitatively reproduce some aspects of a more realistic theory.  These points should be
kept in mind when comparing analysis using (\ref{p-adic}) to the previous analysis using 
(\ref{DBI},\ref{BSFT},\ref{Lambert},\ref{Laidlaw}).  That being said we note that there are several 
nontrivial
qualitative similarities between $p$-adic string theory and tachyon matter.  For example, near the
true vacuum of the theory $\phi=0$ the field naively has no dynamics since its mass squared goes to infinity
\footnote{Reference \cite{Zwiebach:p-adic} found anharmonic oscillations around the vacuum 
by numerically solving the full nonlinear equation of motion. However, these 
solutions do not correspond to conventional physical
states.}.  This is the $p$-adic version of the statement
that there are no open string excitations at the tachyon vacuum.  A second similarity between $p$-adic 
string 
theory and tachyon matter is the existence of lump-like soliton solutions representing $p$-adic 
D-branes \cite{Sen:p-adic}.  The theory of small fluctuations about these lump solutions has a spectrum of 
equally spaced masses squared for the modes \cite{Minahan:ModeInteractions}, as in the case of normal 
bosonic string theory.  On the other hand, there are some important differences between the theory 
(\ref{p-adic}) and (\ref{DBI},\ref{BSFT},\ref{Lambert},\ref{Laidlaw}).  In the case of tachyon matter the 
vacuum is at infinity and the tachyon never reaches this point, whereas in the case of the $p$-adic string
the vacuum is at a finite point in the field configuration space and homogeneous solutions rolling towards
the vacuum typically pass this point without difficulty \cite{Zwiebach:p-adic}.  In fact, the numerical 
studies of \cite{Zwiebach:p-adic} found \emph{no} homogeneous solutions which appeared to correspond to 
tachyon matter (vanishing pressure at late times).  

Keeping in mind that the connection between (\ref{p-adic}) and 
(\ref{DBI},\ref{BSFT},\ref{Lambert},\ref{Laidlaw}) is not entirely clear we proceed to study inhomogeneous
solutions of (\ref{p-adic}) as an example of a theory with an infinite number of derivatives which may have
some qualitative similarities to the string theory tachyon.  At first glance it is not immediately clear how
to proceed since (\ref{p-adicEOM}) is difficult to solve
for profiles with nontrivial dependence on more than one variable.  The simplest solution is to study
small fluctuations about some known time-dependent solution of (\ref{p-adicEOM}).  We are interested in the
dynamics near the vacuum and \cite{Zwiebach:p-adic} found anharmonic homogenous oscillations near $\phi=0$
so one might consider small spatial inhomogeneities about these solutions.  However, these anharmonic
oscillations cannot be found by solving the linearized equation of motion and hence we should not expect to
be able to study small inhomogeneities near the vacuum by linearizing about these oscillators.  One might 
consider the closest analogy to the solutions found in the preceeding sections to be small fluctuations
about a time dependent solution which interpolates between the unstable vacuum $\phi=1$ (or also $\phi=-1$ 
in the case of odd $p$) and the stable vacuum $\phi=0$.  However, it was shown in 
\cite{Zwiebach:p-adic,Vladimirov:Nonlinear} that no such time dependent solution to (\ref{p-adicEOM})
exists  
\footnote{More precisely, it was shown that there exists no homogeneous nonnegative bounded continuous 
solution of (\ref{p-adicEOM}) with $\phi(t \rightarrow -\infty)=1$ and $\phi(t \rightarrow +\infty) = 0$.}.
We therefore will consider inhomogeneous fluctuations about the rapidly increasing solution
\begin{equation}
\label{rolling}
  \phi_0(t) = p^{\frac{1}{2(p-1)}} \exp \left( \frac{1}{2} \frac{p-1}{p \ln p} t^2  \right).
\end{equation}
We stress that this solution does not roll from the unstable maximum of the potential to the true vacuum
of the theory (indeed no such solution appears to exist) and thus the ensuing analysis may be of limited
relevance since the connection between $p$-adic string theory and tachyon matter is unclear.  The 
homogeneous solution (\ref{rolling}) does bear some qualitative similarity to the solutions of 
(\ref{DBI},\ref{BSFT},\ref{Lambert},\ref{Laidlaw}) in the sense that this solution represents the tachyon
rolling down the potential though in the case of (\ref{rolling}) the tachyon rolls towards 
$V \rightarrow -\infty$ and not $V = 0$.

Writing $\phi(t,x) = \phi_0(t) + \delta \phi(t,x)$ with $\phi_0(t)$ given by (\ref{rolling}) the equation
of motion (\ref{p-adicEOM}) is
\begin{equation}
\label{linearizedP-adic}
  p^{-\frac{1}{2} \partial^{\mu} \partial_{\mu}} \delta \phi(t,x) = p \phi_0 (t)^{p-1} \delta \phi(t,x)
\end{equation}
to linear order in $\delta \phi / \phi_0$.   The particular solutions of 
(\ref{linearizedP-adic}) may be written as
\begin{equation}
\label{mode}
  \delta \phi_{\lambda} (t,x) = \phi_0(t) K_{\lambda}( i \alpha t) \chi_{\lambda} (x)
\end{equation}
where
\[
  \alpha = \sqrt{\frac{1}{2}\frac{p^2-1}{p \ln p}}
\]
and $K_{\lambda}(z)$ is any solution of the  Hermite equation
\begin{equation}
\label{HermiteEqn}
  \frac{\partial^2 K_{\lambda}(z)}{\partial z^2}  - 2 z \frac{\partial K_{\lambda}(z)}{\partial z}  + 
  2 \lambda K_{\lambda}(z) =0.
\end{equation}
Plugging (\ref{mode}) into (\ref{linearizedP-adic}) one finds that the spatial modes $\chi_{\lambda} (x)$ 
are determined by the equation
\begin{equation}
\label{modeeqn}
  p^{-\frac{1}{2} \partial_x^2} \chi_{\lambda} (x) = p^{1-\lambda} \chi_{\lambda} (x).
\end{equation}

Before attempting to solve (\ref{modeeqn}) for the spatial dependence of the particular solutions of
(\ref{linearizedP-adic}) some comments are in order concerning the separation of variables (\ref{mode}).
It is straightforward to show that the solutions of (\ref{linearizedP-adic}) separate as (\ref{mode})
using the identity \cite{Minahan:ModeInteractions}
\begin{eqnarray}
\label{identity}
  e^{a \partial_t^2} \left[ K_{\lambda}(i \alpha t) e^{b t^2} \right] 
  &=& (1-4ab)^{-1/2}\left(1 + \frac{4 a \alpha^2}{1-4ab} \right)^{\lambda/2}  \nonumber \\
 &\times& K_{\lambda} \left(\frac{i \alpha t}{\sqrt{(1-4ab)(1-4ab+4a\alpha^2)}} \right) 
 \exp \left(\frac{b t^2}{1-4ab} \right).
\end{eqnarray}
This identity was considered in \cite{Minahan:ModeInteractions} for 
$\lambda = n = 0,1,2,\cdots$ and $K_{n}(z) = H_{n}(z)$, the Hermite polynomials, though it holds for
arbitrary $\lambda$
which is most easily shown by noting that the two solutions of (\ref{HermiteEqn}),
 $K_{\lambda}(z) = P_{\lambda}(z)$ and $K_{\lambda}(z) = Q_{\lambda}(z)$, have contour integral 
representations \cite{Hilbert}
\begin{equation}
\label{P_lambda}
  P_{\lambda} (z) = \int_{C_1} \frac{d \xi}{2 \pi i}  \, \xi^{-\lambda-1} e^{-\xi^2 + 2 \xi z}
\end{equation}
and
\begin{equation}
\label{Q_lambda}
  Q_{\lambda} (z) = \int_{C_2} \frac{d \xi}{2 \pi i}  \, \xi^{-\lambda-1} e^{-\xi^2 + 2 \xi z}
\end{equation}
where the curves $C_1$ and $C_2$ are given in figure \ref{contourFIG1}.  The standard definition of the 
Hermite polynomials for $\lambda = n = 0,1,2,\cdots$ is given by
\begin{eqnarray}
\label{H_n}
  H_{n} (z) &=& n! \left( P_{n} (z) + Q_{n} (z) \right) \nonumber \\
              &=& n! \int_C \frac{d \xi}{2 \pi i}  \, \xi^{-n-1} e^{-\xi^2 + 2 \xi z} 
\end{eqnarray}
where $C$ is the contour given in figure \ref{contourFIG2}.  For subsequent calculations we will be 
particularly interested in defining the Hermite functions $H_{\lambda}(z)$ on $\lambda \leq 1$.
For $\mathrm{Re}(\lambda) < 0$ we can contract the path of integration $C$ to the origin and write
\begin{equation}
\label{lambda<0}
  H_{\lambda} (z) = 
\frac{1}{\Gamma(-\lambda)} \int_0^{\infty}  d\xi \, \xi^{-\lambda -1} e^{-\xi^2 - 2 \xi z} 
\end{equation}
where the normalization has been chosen to agree with \cite{HermiteFunctions}.  Note that when
$\mathrm{Re}(\lambda)>-1$ we can also represent the Hermite functions by a real integral
\cite{HermiteFunctions}
\begin{equation}
\label{lambdaGNegOne}
  H_{\lambda} (z) = \frac{2^{\lambda} e^{z^2}}{\sqrt{\pi}} \int_0^{\infty} e^{-\xi^2} \xi^{\lambda}
                    \cos \left( 2 z \xi - \frac{\lambda \pi}{2} \right)
\end{equation}
which coincides with the standard definition of the Hermite polynomials when $\lambda = n = 0,1,2,\cdots$.
With these conventions the Hermite functions are normalized so that
\[
   H_{\lambda} (0) = \frac{2^{\lambda} \Gamma(1/2)}{\Gamma \left(\frac{1-\lambda}{2}\right)}, \hspace{5mm}
   H_{\lambda}' (0) = \frac{2^{\lambda} \Gamma(-1/2)}{\Gamma \left(-\lambda / 2\right)}.
\]

\vspace{-6mm}

\DOUBLEFIGURE[ht]{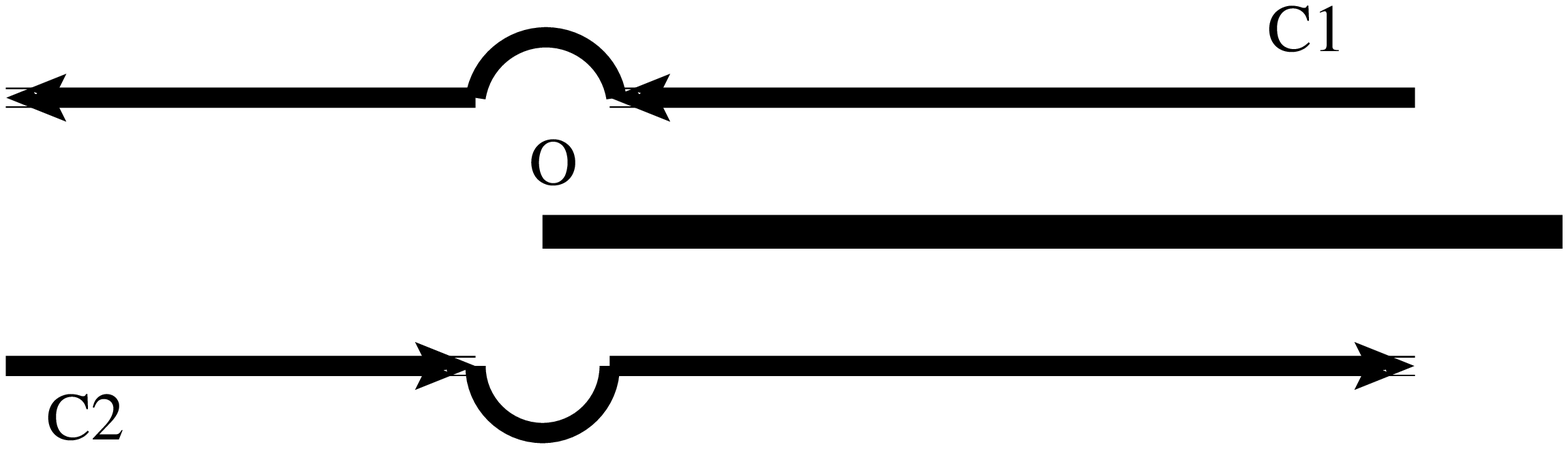,width=2in,height=1.5in}{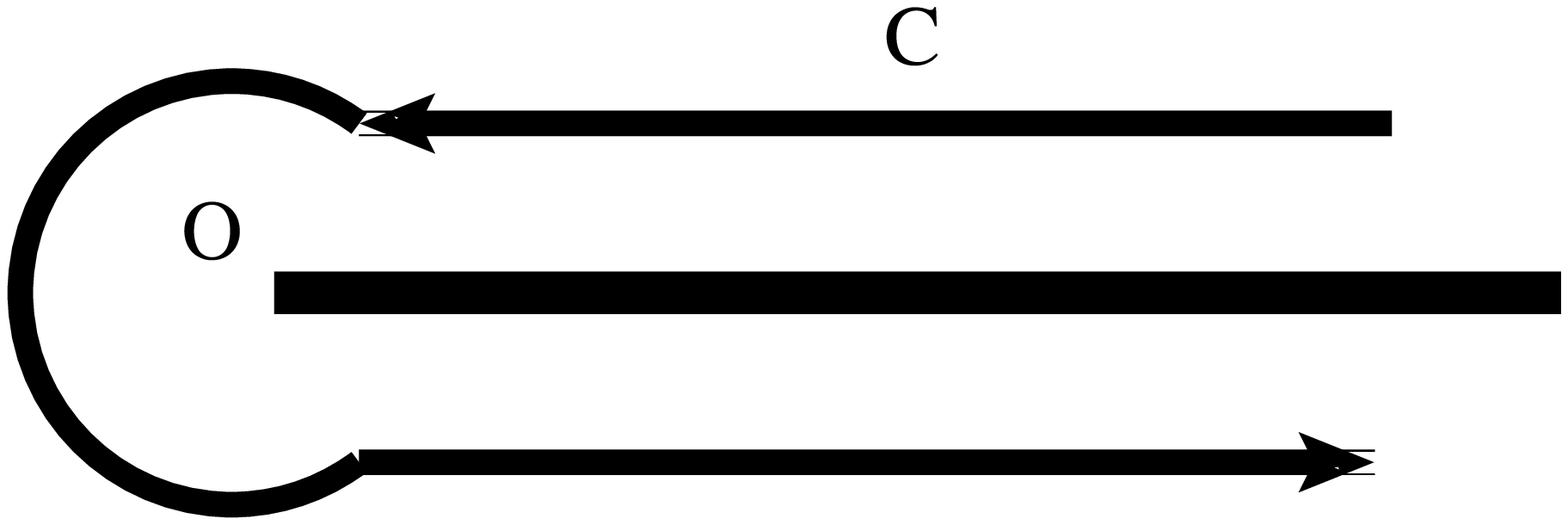,width=2in,height=1.5in}{The contours of integration $C_1$ and $C_2$.\label{contourFIG1}}{The contour of integration $C$. \label{contourFIG2}}

We now proceed to determine the spatial dependence of the modes (\ref{mode}).
Equation (\ref{modeeqn}) has solutions
\begin{equation}
\label{modesoln}
  \chi_{\lambda} (x) = a_{\lambda} \cos(\omega_{\lambda} x) + 
  b_{\lambda} \sin(\omega_{\lambda} x )
\end{equation}
where $\omega_{\lambda} = \sqrt{2(1-\lambda)}$.  For initial data which are periodic on some
interval $\left[-L,+L\right]$ one takes $\omega_{\lambda} = \pi m / L$ where $m=0,1,2,\cdots$.  
With this choice the degree of the Hermite function is
\[
  \lambda = 1 - \frac{\pi^2 m^2}{2 L^2}.
\]
For the zero mode $m=0$ we have $\lambda=1$ and for all other $m$ we have $\lambda \not= 0,1,2,\cdots$
unless $L$ is chosen so that $\sqrt{2} \, L / \pi$ is an integer.

The Hermite equation (\ref{HermiteEqn}) is second order and we expect to be able to find two linearly 
independent solutions for each $\lambda$.  It is most convenient to choose one of these to be the Hermite 
functions as defined by (\ref{H_n},\ref{lambda<0},\ref{lambdaGNegOne}).   To obtain a second solution we 
note 
the Wronskian formula \cite{HermiteFunctions}
\begin{equation}
\label{wronskian}
  W \left[ H_{\lambda} (i z), H_{\lambda} (-i z) \right] = 
  \frac{ 2^{\lambda+1} \sqrt{\pi} }{\Gamma(-\lambda)} e^{-z^2}.
\end{equation}
so that 
\begin{eqnarray*}
  \delta \phi_{\lambda}^{(+)} (t,x) &=& \phi_0(t) H_{\lambda}( +i \alpha t) \, \chi_{\lambda}^{(+)} (x), \\
  \delta \phi_{\lambda}^{(-)} (t,x) &=& \phi_0(t) H_{\lambda}( -i \alpha t) \, \chi_{\lambda}^{(-)} (x) 
\end{eqnarray*}
are linearly independent for $\lambda \not= 0, 1, 2, \cdots$.  The linear independence of these two solutions
fails in the exceptional case $\lambda=1$ for all $L$.  The linear independence of these two solutions
may also fail on other values of $m$ if $\sqrt{2}\, L / \pi$ is an integer.
For simplicity we exclude the case of integer
$ \sqrt{2} \, L / \pi$ from the present analysis and take $\lambda=1$ to be the only special case.
(We note, however, that is is straighforward to extend our analysis to include integer values of
$ \sqrt{2} \, L / \pi$.)  In the case $\lambda=1$ ($m=0$)
we can take, as one solution of (\ref{HermiteEqn}), the Hermite polynomial (\ref{H_n})
\[
  H_1 (z) = 2 z.
\]
The second solution may be written formally as
\[
    P_{1} (z) - Q_{1} (z) 
\]
where the functions $P_{\lambda}$ and $Q_{\lambda}$ are given by (\ref{P_lambda},\ref{Q_lambda}).  The 
function obtained in this manner does not have a  simple closed form expression as does $H_{1}(z)$.

Putting the results of this section together we write the general solutions of (\ref{linearizedP-adic}) as
\begin{eqnarray}
\label{general}
  \delta \phi(t,x) = 
   &\!\!\!\!\!\!\!\!\!\!\!\!\!\!\!\!\!\!\!\!\!\!\!\!\!\!\!\!\!\!\!\!\!\!\!\!\!\!\!\!\!\!\!\!\!\!\!\!\!\!\!\!
    \!\!\!\!\!\!\!\!\! 
    \phi_0(t) \,
    \mathrm{Re} \left[ P_{1} (i \alpha t) - Q_{1} (i \alpha t) \right]\frac{1}{2}\, a_0^{(1)} +
    \phi_0(t) \, \frac{t}{2} \, a_0^{(2)} & \\
   &\phi_0(t) \sum_{m=1}^{\infty} \tilde{H}_{1-\pi^2m^2 / 2 L^2}^{(1)} (t) \, \chi_{m}^{(1)} (x) +
     \phi_0(t) \sum_{m=1}^{\infty} \tilde{H}_{1-\pi^2m^2 / 2 L^2}^{(2)} (t) \, \chi_{m}^{(2)} (x)& 
 \nonumber
\end{eqnarray}
where we have defined the functions
\[
  \tilde{H}^{(1)}_{\lambda}(t) = \frac{\Gamma(\frac{1-\lambda}{2})}{2^{\lambda}\Gamma(1/2)}\frac{1}{2} 
  \left( H_{\lambda}(i \alpha t) + H_{\lambda}(-i \alpha t)  \right)
\]
and
\[
  \tilde{H}^{(2)}_{\lambda}(t) = \frac{\Gamma(-\lambda / 2)}{2^{\lambda}\Gamma(-1/2)}\frac{1}{2 \alpha i} 
  \left( H_{\lambda}(i \alpha t) - H_{\lambda}(-i \alpha t)  \right).
\]
The normalizations have been chosen so that $\tilde{H}^{(1)}_{\lambda}(0)=1$ and 
$\partial_{t} \tilde{H}^{(2)}_{\lambda}(t) |_{t=0} = 1$.  The spatial modes $\{ \chi_m^{(i)} \}$ are
\[
  \chi_m^{(i)}(x) = a_m^{(i)} \cos \left( \frac{\pi m x}{L} \right) + 
  b_m^{(i)} \sin \left( \frac{\pi m x}{L} \right) 
\]
for $i=1,2$.  The coefficients $\{a_m^{(1)},b_m^{(1)}\}$ determine the fourier expansion of 
$\delta \phi / \phi_0$ at $t=0$ and the coefficents $\{a_m^{(2)},b_m^{(2)}\}$ determine the fourier 
expansion of $\partial_t (\delta \phi / \phi_0)$ at $t=0$.

For practical numerical computations it is simplest to consider initial data where $a_0^{(1)}=0$ to eliminate
the function $P_1 (i \alpha t) - Q_1 (i \alpha t)$ from the expression (\ref{general}).  
For example, this will be true for any initial data such that $\delta \phi / \phi_0$ at $t=0$ is an odd 
function of $x$.

Recall that the linearized equation (\ref{linearizedP-adic}) is valid for 
$| \delta \phi(t,x) / \phi_0(t) | \ll 1$.  The stability of the solution (\ref{general}) therefore
depends on the asymptotic behaviour of the functions $\tilde{H}^{(1)}_{\lambda}(t)$ and 
$\tilde{H}^{(2)}_{\lambda}(t)$ at late times.
It is straightforward to show  that $\tilde{H}^{(1)}_{\lambda}(t)$ and 
$\tilde{H}^{(2)}_{\lambda}(t)$ have the asymptotic behaviour
\[
  \tilde{H}^{(1)}_{\lambda}(t) \sim \tilde{H}^{(2)}_{\lambda}(t) \sim t^{\lambda}
\]
for large $t$.  That these two solutions become linearly dependent as $t \rightarrow \infty$ was to be
anticipated from (\ref{wronskian}).  We find, then, that for modes with 
$\lambda = 1 - \frac{\pi^2 m^2}{2 L^2} < 0$ one has 
$\delta \phi_{m}^{(i)} (t,x) / \phi_0(t) \rightarrow 0$ and the linearized approximation is stable.
On the other hand, modes with $\lambda = 1 - \frac{\pi^2 m^2}{2 L^2} > 0$ 
\footnote{Recall that we are excluding the case $\lambda=0$.}
are increasing functions of time and hence the linearized approximation will eventually break down.
In the case that $L < \pi / \sqrt{2} $ and $a_0^{(1)} = a_0^{(2)} = 0$ then all of the  modes are decreasing 
and the perturbative expansion is reliable throughout the evolution.
In this case we can make definitive statements about the absence of caustics or 
similar singularities in the theory (\ref{p-adic}).  In fact, for initial data which satisfy these 
restrictions, the profile $\phi(t,x)$ becomes homogeneous at late times.  
One might argue that (\ref{general}) implies the absence of caustic formation even for more general initial 
data (for example when $L$ is large and many of the modes $\delta \phi_{m}^{(i)} (t,x) / \phi_0(t)$ are 
increasing) since in this case the homogeneous zero mode of $\delta \phi  / \phi_0$ increases faster than 
all other modes.

\vspace{8mm}

\DOUBLEFIGURE[ht]{p-adic3.eps,width=2.5in,height=2in}{p-adic2.eps,width=2.5in,height=2in}{Plot of the spatial profiles $\delta \phi  / \phi_0$ for a series of increasing time steps.  The initial data is such that the linearized approximation is valid throughout the evolution.\label{padicFIG3}}{Plot of the spatial profiles $\delta \phi  / \phi_0$ for a series of increasing time steps.  The initial data is such that the linearized approximation breaks down at late times.  \label{padicFIG4}}

Figure \ref{padicFIG3} shows a plot of the spatial profiles $\delta \phi  / \phi_0$ given by (\ref{general})
for a series of increasing time steps 
for the case $p=2$, $L=1$.  The initial data are 
$\delta \phi (0,x) / \phi_0(0) = 0.01 \, x \, e^{-(x / 0.4)^2}$ and 
$\partial_t (\delta \phi (t,x) / \phi_0(t))|_{t=0} = 0.001 \, \sin(\pi x / L)$
on $-L \leq x \leq +L$.  For this choice of initial data $a_0^{(1)} = a_0^{(2)} = 0$ and 
$\lambda < 0$ for all $m$.  The perturbative expansion is stable for this example and the field tends
towards a homogeneous profile at late times.  Figure \ref{padicFIG4} shows the same plot for the initial
data $\delta \phi (0,x) / \phi_0(0) = 0.01 \, x \, e^{-(x / 0.4)^2}$ and 
$\partial_t (\delta \phi (t,x) / \phi_0(t))|_{t=0} = 0.01$ on $-L \leq x \leq +L$.  With this choice of 
initial data $\delta \phi / \phi_0 \sim t$ at late times since $a_0^{(2)} \not= 0$ and the linearized
approximation will eventually break down.

Finally we comment on the interpretation of the Cauchy problem for the linearized differential equation
(\ref{linearizedP-adic}).  We have found a general solution of (\ref{linearizedP-adic}) for which we are
free to specify the initial field $\delta \phi / \phi_0$ at $t=0$ and the time derivative
$\partial_t (\delta \phi / \phi_0)$ at $t=0$.  This may seem surprising since (\ref{linearizedP-adic})
contains an infinite number of time derivatives and one might naively expect to have the freedom to specify
an infinite number of initial data $\partial_t^{(n)} (\delta \phi / \phi_0)|_{t=0}$ for $n=0,1,2,\cdots$.
However, the initial value problem for homogeneous solutions of (\ref{p-adicEOM}) was studied in 
\cite{Zwiebach:p-adic} and it was found that the equation of motion itself imposes an infinite number of 
consistency conditions on the initial data which one can consider.  In fact, it was speculated in 
\cite{Zwiebach:p-adic} that the space of allowable initial conditions for (\ref{p-adicEOM}) may be finite.
This conjecture seems consistent with our results.  It is interesting to note that (\ref{linearizedP-adic})
seems to be an example of an equation containing an infinite number of derivatives whose solution space is 
surprisingly similar to that of equations containing only two time derivatives.

\section{Conclusions}

We have studied several different effective actions describing the open superstring tachyon.  In the case
of actions containing only first derivatives of the tachyon field we have studied three different theories 
proposed on different grounds.  These three actions are expected to be valid in different limits of
string theory, however, we find that that 
vacuum structure of all three theories is remarkably similar.  In particular, we have shown that all three
theories lead to the formation of caustics where the field becomes multi-valued and where second and higher 
derivatives of the field blow up.  Each of these three actions cannot be trusted in the vicinity of a 
caustic since 
higher derivative corrections which have been neglected become important.  We considered also an effective 
action containing second order derivatives of the field and found a similar structure of derivative 
singularities.  Finally, in the context of $p$-adic string theory, we studied small inhomogeneities about a 
time dependent solution which rolls from the unstable vacuum to infinity in field configuration space.  
In this case we found that for a broad class of initial data the linearized approximation is reliable
throughout the evolution and we could conclusively show the absence of caustics or similar pathologies.
This result seems suggestive that the formation of caustics is an artifact of truncating a theory with an
infinite number of derivatives, however, the connection between $p$-adic string theory and tachyon matter is
still unclear. 

Since we have restricted our analysis to the study of effective actions we cannot conclude if the phenomenon
of caustic formation is a genuine prediction of string theory or not.  If this prediction is borne out by 
string theory it will be necessary to find some physical interpretation of the caustics, though none is 
obvious to us.  As pointed out in \cite{Caustics} such an interpretation could depend on the dimensionality 
of the theory.  If these caustics are in fact a real prediction of string theory it will also be necessary 
to find some way to predict the field value in the vicinity of a caustic.  The situation is qualitatively 
similar to the formation of shock waves in nonlinear gas dynamics.  In that case the multi-valued 
continuous solution in the vicinity of a shock is replaced with a single valued discontinuous solution using
the Rankine-Hugoniot jump condition to quantify the discontinuity in the field.  It is possible that some 
similar auxiliary condition could be used to predict the value of the tachyon field in the vicinity of a 
caustic. 

\acknowledgments

The author would like to thank M. Laidlaw for useful correspondence and J. Cline for many helpful 
discussions and comments on the manuscript.  This research is supported by the Natural Sciences and 
Engineering Research Council (NSERC).

\renewcommand{\theequation}{A-\arabic{equation}}
\setcounter{equation}{0}  

\section*{APPENDIX: The Legendre Transformation}

In $1+1$ dimensions equation (\ref{leadinghigher}) takes the form
\begin{equation}
\label{1+1}
  \dot{T}^2 \ddot{T} + T'^2T'' - 2 \dot{T}T' \dot{T}' = 0
\end{equation}
To construct general solutions of (\ref{1+1}) we preform the Legendre transformation (see \cite{Handbook} 
for example) defined by the relations
\begin{equation}
\label{legendre}
  \xi = T', \hspace{5mm}
  \eta = \dot{T}, \hspace{5mm}
  T(t,x) + \omega(\eta,\xi) = x \xi + t \eta
\end{equation}
where $\omega(\eta,\xi)$ is a new dependent variable and $\{\eta,\xi\}$ are the new independent variables.
It follows from (\ref{legendre}) that
\begin{equation}
\label{x,t}
  x = \frac{\partial \omega}{\partial \xi}, \hspace{5mm} t = \frac{\partial \omega}{\partial \eta}.
\end{equation}
It is straightforward to verify the relations
\begin{equation}
\label{secondderivatives} 
  T'' = J \frac{\partial^2 \omega}{\partial \eta^2}, \hspace{5mm}
  \ddot{T} = J \frac{\partial^2 \omega}{\partial \xi^2}, \hspace{5mm}
  \dot{T}' = -J \frac{\partial^2 \omega}{\partial \xi \partial \eta} 
\end{equation}
where the Jacobian of the transformation is 
\begin{eqnarray}
\label{jacobian}
  J &=& T'' \ddot{T} - \left( \dot{T}' \right)^2 \\ \nonumber
    &=& \left[ \frac{\partial^2 \omega}{\partial \xi^2} \frac{\partial^2 \omega}{\partial \eta^2} -
      \left( \frac{\partial^2 \omega}{\partial \xi \partial \eta} \right)^2 \right]^{-1}
\end{eqnarray}
which we assume is nonzero.  Note that this analysis excludes any solutions where $J=0$.
The nonlinear partial differential equation (\ref{1+1}) is transformed to a linear partial differential 
equation in terms of the new variables
\begin{equation}
\label{omegaPDE}
    \xi^2 \frac{\partial^2 \omega}{\partial \eta^2} 
  + \eta^2 \frac{\partial^2 \omega}{\partial \xi^2}
  + 2 \eta \xi \frac{\partial^2 \omega}{\partial \xi \partial \eta} =0.
\end{equation}
Given a solution $\omega(\eta,\xi)$ of (\ref{omegaPDE}) the relations (\ref{x,t}) along with
\begin{equation}
\label{invert}
  T = \xi \frac{\partial \omega}{\partial \xi} + \eta \frac{\partial \omega}{\partial \eta} 
        -\omega(\eta,\xi)   
\end{equation}
define the solution of (\ref{1+1}) parametrically.

We consider now solutions of the linear partial differential equation (\ref{omegaPDE}).
To simplify the ensuing analysis we will restrict ourselves to solutions where 
$\partial_{\mu} T \partial^{\mu} T$ does not change sign.  This is expected to be a reasonable restriction 
for the vacuum structure of the theory (\ref{Laidlaw}) since previous analysis of the actions 
(\ref{DBI},\ref{BSFT},\ref{Lambert}) suggests that $\partial_{\mu} T \partial^{\mu} T$ is either increasing
or decreasing as $T \rightarrow \pm \infty$.  We only consider the case 
$\partial_{\mu} T \partial^{\mu} T \leq 0$, which we expect to most closely 
resemble  the vacuum  structure of the actions (\ref{DBI},\ref{BSFT},\ref{Lambert}).  In this regime we
can define new coordinates $\{\rho,\sigma\}$ by
\begin{equation}
\label{newcoords1}
  \eta = \rho \cosh \sigma, \hspace{5mm} \xi = \rho \sinh \sigma
\end{equation}
such that
\[
   \rho^2 = -\xi^2 + \eta^2 = -\partial_{\mu} T \partial^{\mu} T \geq 0, \hspace{5mm}
   \tanh \sigma = \frac{\xi}{\eta}.
\]
In terms of these new variables (\ref{omegaPDE}) takes the remarkably simple form
\begin{equation}
\label{newPDE}
  - \rho \frac{\partial \omega}{\partial \rho} + \frac{\partial^2 \omega}{\partial \sigma^2} =0.
\end{equation}
with particular solution
\[
  \omega_m (\eta,\xi) = \rho^{m^2} \left[ \alpha_m e^{m \sigma} + \beta_m e^{-m \sigma} \right] 
\]
for arbitrary $m$, $\alpha_m$, $\beta_m$.  The most convenient way to fix the Cauchy data is to specify 
$\omega$ and
$\partial \omega / \partial \rho$ at $\sigma = 0$.  In this case we write the general solution of 
(\ref{newPDE}) as
\begin{equation}
\label{generalomega}
  \omega (\rho,\sigma) = \sum_{n=0}^{\infty} \rho^n \left( a_n \cosh \left( \sqrt{n} \, \sigma \right) +
          b_n \sinh \left( \sqrt{n} \, \sigma \right)     \right)
\end{equation}
where $\{a_n\}$ determine the coefficients in the taylor expansion of $\omega$ at $\sigma=0$ and 
$\{b_n\}$ determine the coefficients in the taylor expansion of $\partial \omega / \partial \rho$ at 
$\sigma = 0$.

It is straightforward to compute $x(\eta,\xi)$, $t(\eta,\xi)$ from (\ref{x,t}) and
\[
  T(\rho,\sigma) = \sum_{n=0}^{\infty} \rho^n (n-1)\left( a_n \cosh \left( \sqrt{n} \, \sigma \right) +
          b_n \sinh \left( \sqrt{n} \, \sigma \right)     \right)
\]
from (\ref{invert}).  Because this solution is not in closed form and defined parametrically, it is difficult
to get an intuitive feeling for the behaviour of $T$ as a function of $t$ and $x$.  However, we believe that 
these general solutions do admit derivative singularities and have, for several choices of $\{a_n,b_n\}$,
found points $(\rho_c,\sigma_c)$ corresponding to finite $x$ and $t>0$ at which $T''$ is singular though 
$T$, $T'$ and $\dot{T}$ are regular.  In this case the second derivative blows up because the Jacobian 
of the Legendre transformation (\ref{jacobian}) is singular.

\end{document}